\documentclass[aps,prb,twocolumn,showpacs,groupedaddress, amsmath]{revtex4}

\usepackage{graphicx}
\usepackage{amssymb}

\begin{document}


\title{Observation of a magnetic-field-induced Griffiths phase in three-dimensional Ising random magnet Ni$_{p}$Mg$_{1-p}$(OH)$_{2}$ from absorption of AC magnetic susceptibility}


\author{Masatsugu Suzuki}
\email[]{suzuki@binghamton.edu}
\affiliation{Department of Physics, State University of New York at Binghamton, Binghamton, New York 13902-6000}

\author{Itsuko S. Suzuki}
\email[]{itsuko@binghamton.edu}
\affiliation{Department of Physics, State University of New York at Binghamton, Binghamton, New York 13902-6000}


\date{\today}

\begin{abstract}
The nature of the Griffiths phase in three-dimensional Ising random magnet Ni$_{p}$Mg$_{1-p}$(OH)$_{2}$ ($p$ = 0.10, 0.25, 0.315, 0.50, 0.80, and 1) is studied from the measurements of absorption $\chi^{\prime\prime}$ (the out-of phase in AC magnetic susceptibility) in the field-cooled (FC) state. The temperature dependence of $\chi^{\prime\prime}$ is measured in the vicinity of the critical temperature [N\'{e}el temperature for $p$ = 0.50, 0.80, 1, and the spin glass transition temperature $T_{SG}$ for $p$ = 0.1, 0.25, 0.315] in the presence of an external magnetic field $H$. For $p$ = 0.50, 0.80, and 1, a peak of $\chi^{\prime\prime}$ vs $T$ due to the metamagnetic transition drastically shifts to the low temperature side with increasing $H$, while a broad peak of $\chi^{\prime\prime}$ vs $T$ is clearly observed well above the N\'{e}el temperature $T_{N}(H=0)$ for $H>20$ kOe. This result gives a piece of evidence for the magnetic-field induced Griffiths phase, where the Griffiths temperature $T_{G}$ is defined as the peak temperature of $\chi^{\prime\prime}$ vs $T$ above $T_{N}(H=0)$. This $T_{G}$ shifts to the low temperature side with increasing $H$, reflecting the nature of antiferromagnetic fluctuations. The peak height $\chi^{\prime\prime}(T_{G})$ drastically increases with increasing $H$ for $H>20$ kOe, following a power form, $\chi^{\prime\prime}(T_{G}) \approx H^{m-1}$. The exponent $m$ is $3.12 \pm 0.04$ for $p = 1$, $2.96 \pm 0.03$ for $p = 0.80$, and $2.47 \pm 0.06$ for $p = 0.50$. For $p$ = 0.10, 0.25, and 0.315, a broad peak of $\chi^{\prime\prime}$ vs $T$ is clearly observed well above $T_{SG}(H=0)$ for $H>20$ kOe. The Griffiths temperature $T_{G}$ shifts to the high temperature side with increasing $H$, reflecting the nature of ferromagnetic fluctuations. The peak height $\chi^{\prime\prime}(T_{G})$ drastically increases with increasing $H$ for $H>20$ kOe, following a power form, $\chi^{\prime\prime}(T_{G})\approx H^{m-1}$. The exponent $m$ is $2.28 \pm 0.08$ for $p = 0.315$, $2.20 \pm 0.07$ for $p$ = 0.25, and $2.35 \pm 0.07$ for $p$ = 0.10.
\end{abstract}


\maketitle



\section{\label{intro}INTRODUCTION}
The observation of the Griffiths phase has been reported in site-diluted Ising systems with a concentration ($p$) of magnetic ions which has a percolation threshold $p_{c}$.\cite{ref01,ref02,ref03,ref04} The Griffiths phase refers to a temperature region between a critical temperature $T_{c}(p)$ associated with the long range spin order and a so-called Griffiths temperature $T_{G}$ [$>T_{c}(c)$].\cite{ref05} The nature of the Griffiths phase is characterized by a statistically rare, but large clusters of unfrustrated spins over the whole system. Such clusters become quasi-ordered below $T_{G}$ and flips collectively with anomalously long characteristic times. The consequence of such processes is slow dynamics of the spin autocorrelation functions $q(t)$ ($=\lbrack \langle S_{i}(0)S_{i}(t)\rangle \rbrack$), whose asymptotic form at large time $t$ is given by an enhanced power law 
\begin{equation} 
q(t)\approx \exp \lbrack -A(\ln t)^{d/(d-1)} \rbrack ,
\label{eq01} 
\end{equation} 
for the Ising systems with short-ranged interaction, where $A$ is a constant, $d$ is the spatial dimension and $\langle\cdots\rangle$ and $\lbrack\cdots\rbrack$ in the expression of $q(t)$ represent thermal and disorder averages respectively. The picture of this cluster dynamics can be natively extended to spin glasses by redefining a cluster as a group of spins connected with few frustrations.\cite{ref05,ref06,ref07,ref08}

In this paper we report the observation of the Griffiths phase in a diluted Ising random magnet Ni$_{p}$Mg$_{1-p}$(OH)$_{2}$ (the Ni concentration $p$)\cite{ref09,ref10,ref11,ref12,ref13,ref14,ref15,ref16} using the absorption $\chi^{\prime\prime}$ of the AC magnetic susceptibility. The sample in the present work was the same as that used in the previous papers (powdered samples).\cite{ref14,ref15} We show that the Griffiths phase is strongly enhanced by the application of magnetic fields for $0.1\le p\le 1$. The absorption $\chi^{\prime\prime}$ vs $T$ shows a broad peak centered at $T_{G}(H)$. The peak height of $\chi^{\prime\prime}$ drastically increases with increasing $H$, forming the magnetic-field-induced Griffiths phase. For $0.50 \le p \le 1$ the system exhibits a metamagnetic transition at high magnetic field. The metamagnetic transition temperature $T_{N}(H)$ greatly decreases with increasing the magnetic field (typically $H>20$ kOe), while the Griffiths temperature $T_{G}(H)$ slightly decreases with increasing $H$. For $0.10\le p\le 0.315$, the metamagnetic behavior no longer exists. In turn a spin glass phase newly appears. There is a phase boundary between this spin glass phase (lower temperature phase) and the Griffiths phase (higher temperature phase). The Griffiths phase is strongly enhanced by the application of $H$. The Griffith temperature $T_{G}$ slightly increases with increasing $H$. 

\section{\label{back}Background}
In a Ni(OH)$_{2}$, Ni$^{2+}$ ions form a triangular lattice with a lattice constant $a=3.117 \AA$ in the $c$ plane. The separation distance between adjacent Ni$^{2+}$ layers is $c=4.595 \AA$. The antiferromagnetic interplanar exchange interactions ($J_{2}=-0.28$ K and $J_{3}=-0.09$ K) are weaker than the ferromagnetic intraplanar exchange interaction ($J_{1}=2.7$ K). Because of a negative uniaxial single ion anisotropy $D$ ($=-0.8$ K), magnetic moments of Ni$^{2+}$ spin ($S=1$) are directed along the $c$ axis. Ni(OH)$_{2}$ undergoes an antiferromagnetic phase transition at the N\'{e}el temperature $T_{N}$ (= 26.5 K). Below $T_{N}$ the 2D ferromagnetic long range order is established in each Ni$^{2+}$ layer. Such 2D ferromagnetic layers are antiferromagnetically coupled along the $c$ axis, forming a three-dimensional (3D) antiferromagnetic order. The numbers of neighbors coupled by the interactions $J_{1}$, $J_{2}$, and $J_{3}$, are denoted by $z_{1}$ (= 6), $z_{2}$ (= 2), and $z_{3}$ (= 12), respectively. The crystal structure and the definition of $J_{1}$, $J_{2}$, $J_{3}$ are given in our previous papers.\cite{ref14,ref15} 

Ni$_{p}$Mg$_{1-p}$(OH)$_{2}$ is a diluted 3D Ising random magnet, where a part of Ni$^{2+}$ ions are randomly replaced by nonmagnetic Mg$^{2+}$ ions.\cite{ref09,ref10,ref11,ref12,ref13,ref14,ref15} The N\'{e}el temperature $T_{N}$ decreases with decreasing concentration $p$ for $p>0.4$. The percolation threshold $p_{c}$ where the critical temperature reduces to zero, is about $p_{c}=0.1$. The curve of $T_{N}$ against $p$ shows a terraced form in the concentration region below $p=0.4$: $T_{N}$ is extrapolated to zero near $p_{1}=0.33$. The value of $p_{1}=1/3$ coincides with the estimated critical concentration when only the first neighbor interaction is taken into account; $p_{1}=2/z_{1}=1/3$. This value of $p_{c}$ (= 0.1) coincides with the estimated critical concentration when the first neighbor interaction is taken into account; $p_{c}=2/(z_{1}+z_{2}+z_{3})=1/10$. As will be discussed in Sec.~\ref{dis},\cite{ref15} the magnetic phase diagram of Ni$_{p}$Mg$_{1-p}$(OH)$_{2}$ consists of spin glass (SG) phase and Griffiths phase for $0.1\le p\le 0.315$, reentrant spin glass (RSG) phase, antiferromagnetic (AF) phase, and the Griffiths phase for $0.5\le p\le 0.8$, and the AF phase and the Griffiths phase for $p=p_{c}=0.1$. 

\section{\label{exp}Experimental procedure}
\subsection{\label{expA}ZFC and FC magnetic susceptibility at $H=1$ Oe}
Before setting up a sample at 298K, a remnant magnetic field in a superconducting magnet was reduced to one less than 3 mOe using an ultra low field capability option of the SQUID (superconducting quantum interference device) magnetometer (MPMS XL-5, Quantum Design). For convenience, hereafter this remnant field is denoted as the state of $H=0$. The sample was cooled from 298 to 1.9 K at $H=0$. Then an external magnetic field $H$ (= 1 Oe) was applied at 1.9 K. A zero-field cooled magnetization ($M_{ZFC}$) was measured with increasing $T$ from 1.9 to 50 K. After the ZFC measurement, the sample was heated and kept at 100 K for 20 min. It was again cooled to 30 K in the presence of the same $H$. A field-cooled magnetization ($M_{FC}$) was measured with decreasing $T$ from 30 to 1.9 K. The ZFC and FC magnetic susceptibilities are defined as $\chi_{ZFC}=M_{ZFC}/H$ and $\chi_{FC}=M_{FC}/H$. 

\subsection{\label{expB}Measurement of the AC magnetic susceptibility in the FC state}
The $T$ dependence of the dispersion $\chi^{\prime}$ and the absorption $\chi^{\prime\prime}$ was measured in the presence of various magnetic field $H$, where the amplitude of AC magnetic field $h$ is typically 2 or 3 Oe and the frequency is $f$ = 1 Hz. The sample was cooled from 298 to 1.9K at $H$ = 0. An external magnetic field (0$<H<50$ kOe) was applied at 1.9K. Then $\chi^{\prime}$ and $\chi^{\prime\prime}$ were simultaneously measured with increasing temperature $T$ from 1.9 to 30 K in the presence of fixed $H$. After each $T$ scan, $H$ was changed at 30K and $T$ was decreased from 30 to 1.9 K. Then the measurement was repeated at this $H$. This means that the AC susceptibility was measured in the FC state. 

\section{\label{result}RESULT}
\subsection{\label{resultA}Magnetic-field-induced Griffiths phase in $p=1$}

\begin{figure}
\includegraphics[width=7.0cm]{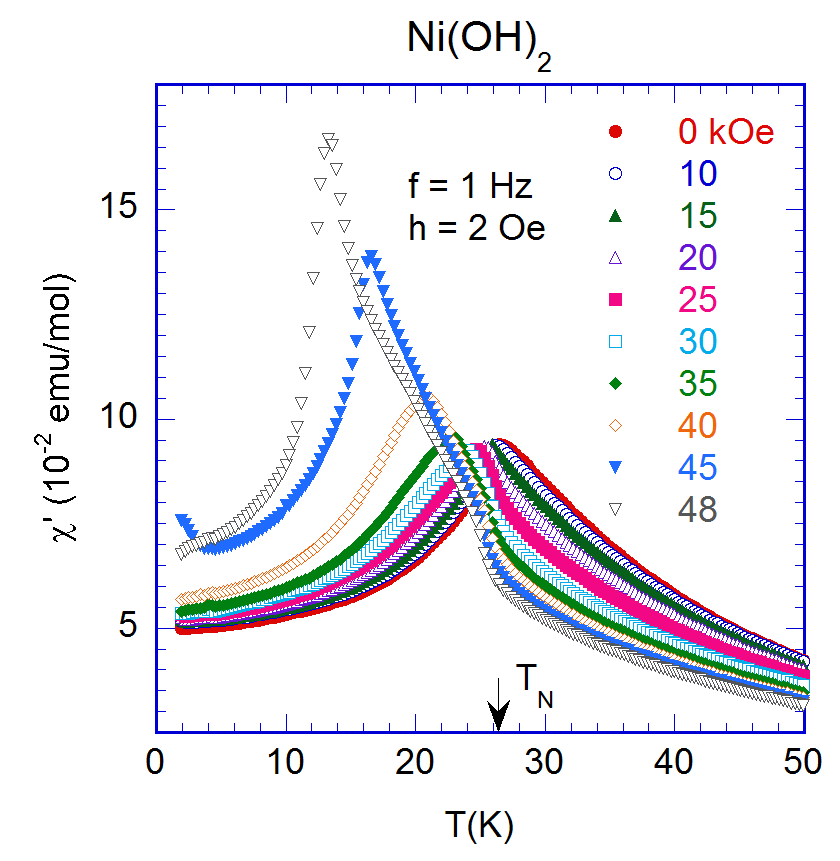}
\caption{\label{fig01}(Color online) $T$ dependence of $\chi^{\prime}$ for $p=1$ at various magnetic fields. $H$ = 0, 10, 15, 20, 25, 30, 35, 40, 45, and 48 kOe. $T_{N}(H=0) = 26.5$ K (denoted by arrow). $f=1$ Hz. $h=2$ Oe. Each measurement of $\chi^{\prime}$ vs $T$ was carried out in the FC state.}
\end{figure}

\begin{figure}
\includegraphics[width=7.0cm]{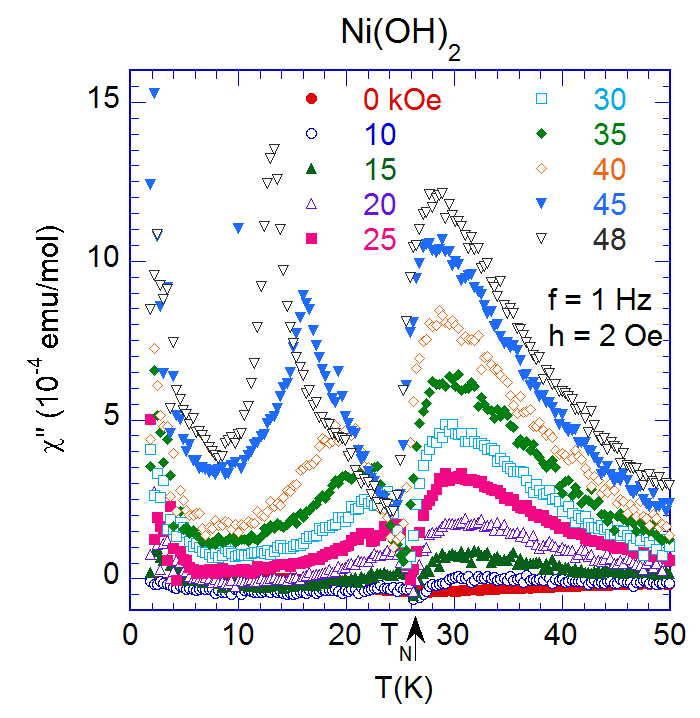}
\caption{\label{fig02}(Color online) $T$ dependence of $\chi^{\prime\prime}$ for $p=1$ at various magnetic fields. $H$ = 0, 10, 15, 20, 25, 30, 35, 40, 45, and 48 kOe. $f=1$ Hz. $h=2$ Oe. Each measurement of $\chi^{\prime\prime}$ vs $T$ was carried out in the FC state. $T_{N}(H=0)=26.5$ K (denoted by arrow). Negative value of $\chi^{\prime\prime}$ arises from the experimental uncertainty.}
\end{figure}

\begin{figure}
\includegraphics[width=7.0cm]{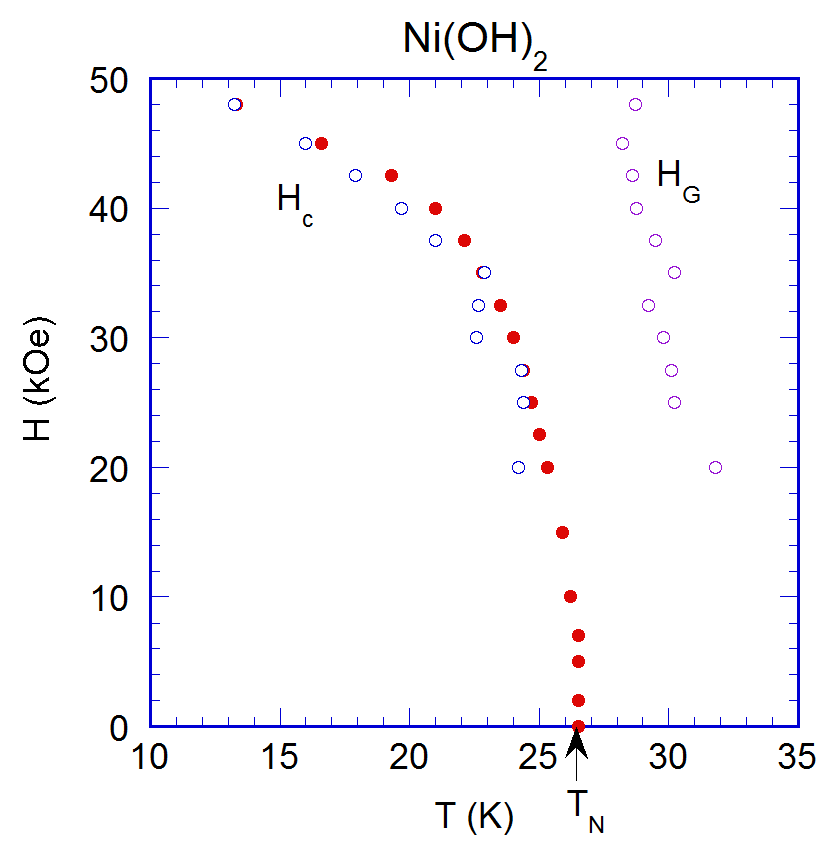}
\caption{\label{fig03}(Color online) Magnetic $H$-$T$ phase diagram of Ni(OH)$_{2}$: peak temperatures of $\chi^{\prime}$ (closed circles) and $\chi^{\prime\prime}$ (open circles). The $H_{G}$ line denotes the phase boundary between the magnetic-field-induced Griffiths phase and the PM phase. The $H_{c}$ line denotes the phase boundary between the AF phase and the magnetic-field-induced Griffiths phase. The critical field is $H_{c} = 55 \pm 1$ kOe at $T$ = 1.56 K (Enoki et al.\cite{ref12}). The N\'{e}el temperature $T_{N}(H=0)=26.5$ K (denoted by arrow).}
\end{figure}

\begin{figure}
\includegraphics[width=7.0cm]{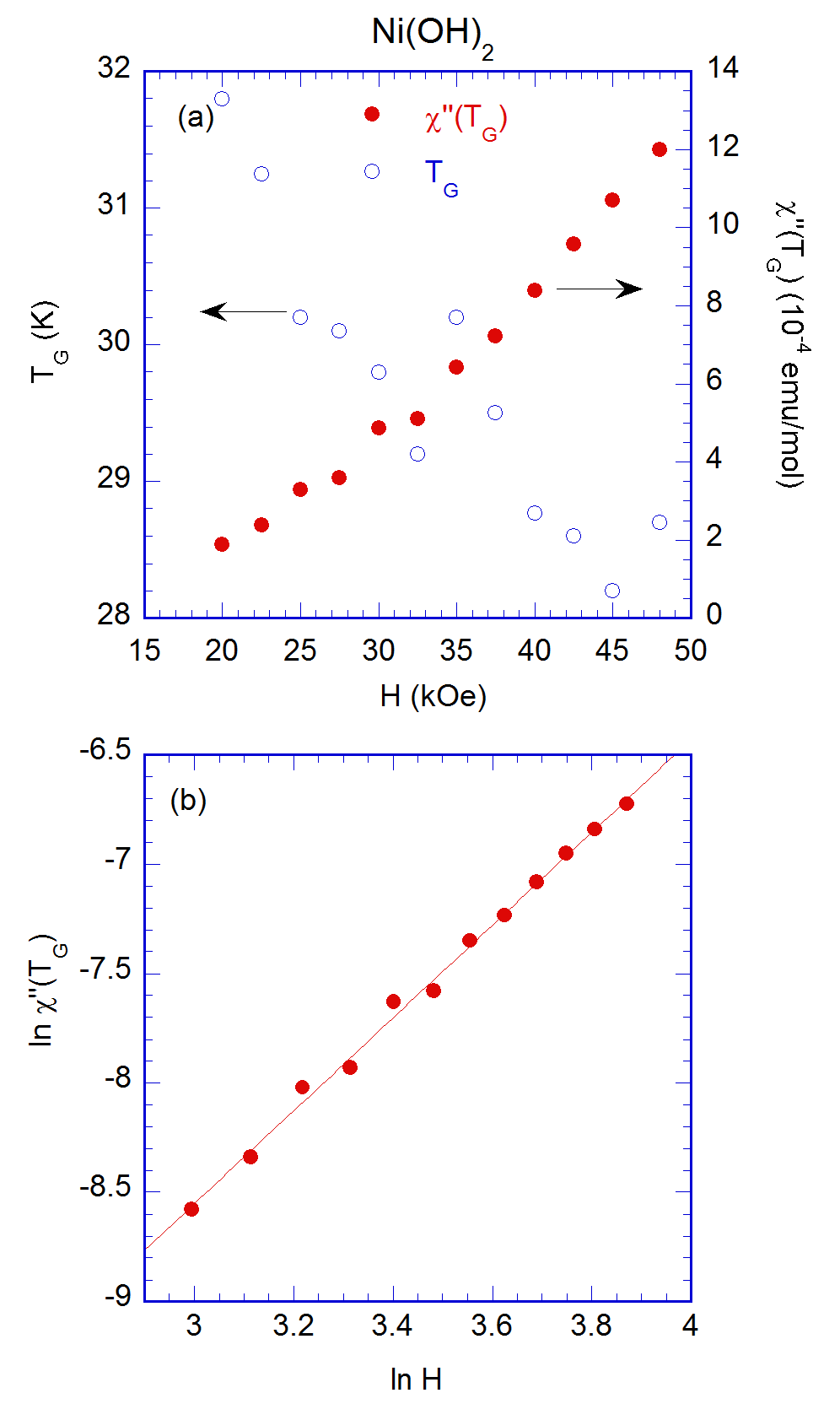}
\caption{\label{fig04}(Color online) (a) Plot of the Griffiths temperature $T_{G}$ as a function of the magnetic field $H$ for $p=1$. The Griffiths temperature $T_{G}$ is defined as a temperature at which $\chi^{\prime\prime}$ vs $T$ has a peak for each $H$. $T_{G}$ decreases with increasing $H$. Plot of the peak height $\chi^{\prime\prime}$($T=T_{G}$) as a function of $H$. The peak height drastically increases with increasing $H$ for $H>20$ kOe. (b) Plot of $\ln \chi^{\prime\prime}(T_{G})$ vs $\ln H$, where $H$ is in the units of kOe and $\chi^{\prime\prime}(T_{G})$ is the units of emu/mol. The least-squares fitting curve is denoted by a straight line.}
\end{figure}

In our previous paper\cite{ref15} we have reported the temperature dependence of the ZFC and FC susceptibilities $\chi_{ZFC}$ and $\chi_{FC}$ for Ni(OH)$_{2}$ in the presence of $H$ = 1 Oe. It is found that the difference $\delta$ ($= \chi_{FC}-\chi_{ZFC}$), appears below a characteristic temperature, and increases with decreasing $T$. Here we define this temperature as the N\'{e}el temperature $T_{N}$ (= 26.5 K). This system shows a metamagnetic behavior in the presence of an external magnetic field. The critical field for the metamagnetic transition is $H_{c}=55 \pm 1$ kOe at $T$ = 1.56 K (Enoki et al.\cite{ref12}). 

Figure \ref{fig01} shows the $T$ dependence of the dispersion $\chi^{\prime}$ in the FC state for $p = 1$ at various magnetic fields. The dispersion shows a sharp peak at $T_{N}$ at $H=0$. No anomaly associated with the Griffiths phase is clearly observed near $T_{N}$. The shift of the peak of $\chi^{\prime}$ to the low temperature side with increasing $H$ is due to the metamagnetic transition between the AF phase and the PM phase (exactly the Griffiths phase). The metamagnetic transition temperature $T_{N}(H)$ drastically decreases with increasing $H$ for $H>35$ kOe. 

Figure \ref{fig02} shows the $T$ dependence of the absorption $\chi^{\prime\prime}$ for $p = 1$ at various magnetic fields, where $f = 1$ Hz. A peak due to the metamagnetic transition drastically shifts to the low temperature side with increasing $H$. No peak is observed around $T=T_{N}$ at $H = 0$. With increasing $H$, a broad peak of $\chi^{\prime\prime}$ vs $T$ starts to appear just above $T_{N}$. This peak slightly shifts to the low temperature side, while the peak height drastically increases with increasing $H$. This broad peak may be associated with the transition between the Griffithts phase and the PM phase. The separation between the peak due to the metamagnetic transition and the peak due to the appearance of the Griffiths phase becomes more prominent for $H>35$ kOe. 

Figure \ref{fig03} shows the magnetic $H$-$T$ diagram for $p=1$. There are two lines, the $H_{G}$ line and $H_{c}$ line. The line $H_{G}$ is the phase boundary between the Griffiths phase and the PM phase, while the line $H_{c}$ is the phase boundary between the AF phase and the Griffiths phase. The critical field is $H_{c} = 55 \pm 1$ kOe at $T = 1.56$ K (Enoki et al.\cite{ref12}). 

Figure \ref{fig04}(a) shows the $H$ dependence of $T_{G}(H)$ for $p = 1$. Here we define $T_{G}(H)$ as a peak temperature of $\chi^{\prime\prime}$ vs $T$. This Griffiths temperature tends to decrease with increasing $H$, for example, $T_{G}$($H$ = 20 kOe) = 31.8 K and $T_{G}$($H$ = 48 kOe) = 28.7 K, reflecting the nature of the antiferromagnetic fluctuations. Note that $T_{G}(H)$ is relatively higher than the N\'{e}el temperature $T_{N}=26.5$ K at $H=0$ for $p=1$. In Fig.~\ref{fig04}(a) we also  show the $H$ dependence of the peak height $\chi^{\prime\prime}(T_{G})$ for $p=1$. This peak height is very small for $H<20$ kOe. It drastically increases with increasing $H$ above 20 kOe. The least squares fit of the $\chi^{\prime\prime}(T_{G})$ vs $H$ (\ref{fig04}(b)) to 
\begin{equation} 
\ln \chi^{\prime\prime}(T_{G}) = A+\zeta \ln H ,
\label{eq02} 
\end{equation} 
for $20<H<48$ kOe, yields the values of $A$ and $\zeta$ as 
\[
A = -14.291 \pm 0.15,\;\;\; \zeta = 2.12 \pm 0.04,
\]
where $A$ and $\zeta$ are the fitted parameter, $H$ is in the units of kOe, and $\chi^{\prime\prime}(T_{G})$ is in the units of emu/mol. As will be introduced in Sec.~\ref{dis}, the exponent $m$ for the domain-wall magnetization $M_{wall}$ is related to the exponent $\zeta$ as,
\begin{equation}
m = \zeta + 1 = 3.12 \pm 0.04.
\label{eq03}
\end{equation}

\subsection{\label{resultB}Absorption $\chi^{\prime\prime}$ at $p=0.80$ in the presence of magnetic field}

\begin{figure}
\includegraphics[width=7.0cm]{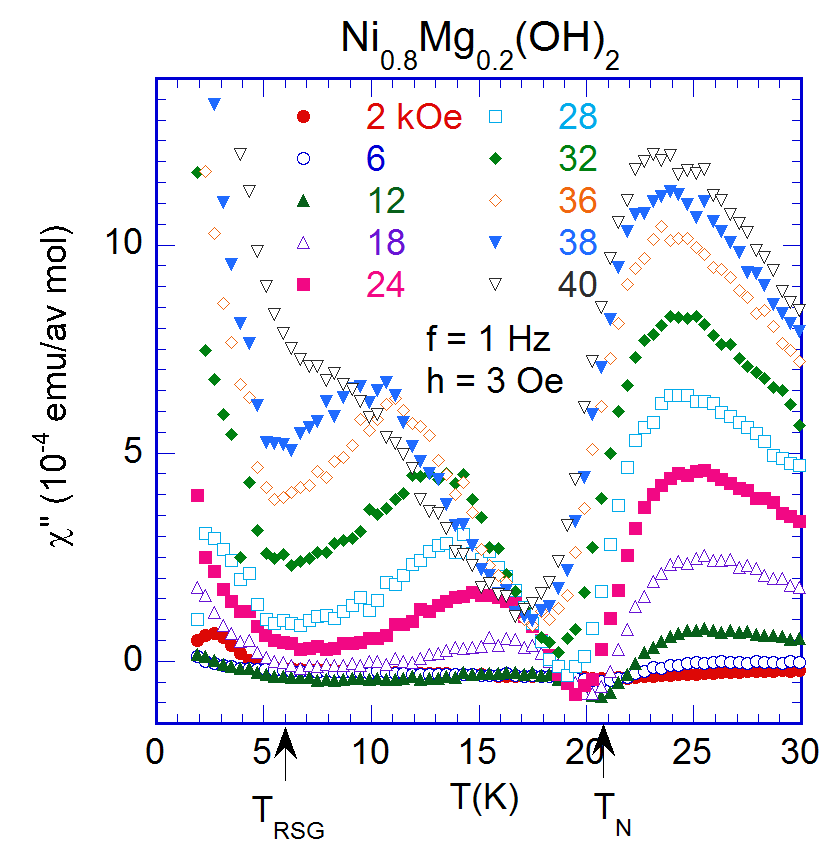}
\caption{\label{fig05}(Color online) $T$ dependence of $\chi^{\prime\prime}$ for $p$ = 0.8 at various magnetic fields. $H$ = 2, 6, 12, 18, 24, 28, 32, 36, 38, and 40 kOe. The reentrant spin glass transition $T_{RSG} = 6$ K and the N\'{e}el temperature $T_{N}(H=0) = 20.7$ K\cite{ref14,ref15} (denoted by arrows). $H_{c} = 44 \pm 1$ kOe at $T = 1.56$ K (Enoki et al.\cite{ref12}). $f = 1$ Hz. $h = 3$ Oe. Each measurement of $\chi^{\prime\prime}$ vs $T$ was carried out in the FC state.}
\end{figure}

\begin{figure}
\includegraphics[width=7.0cm]{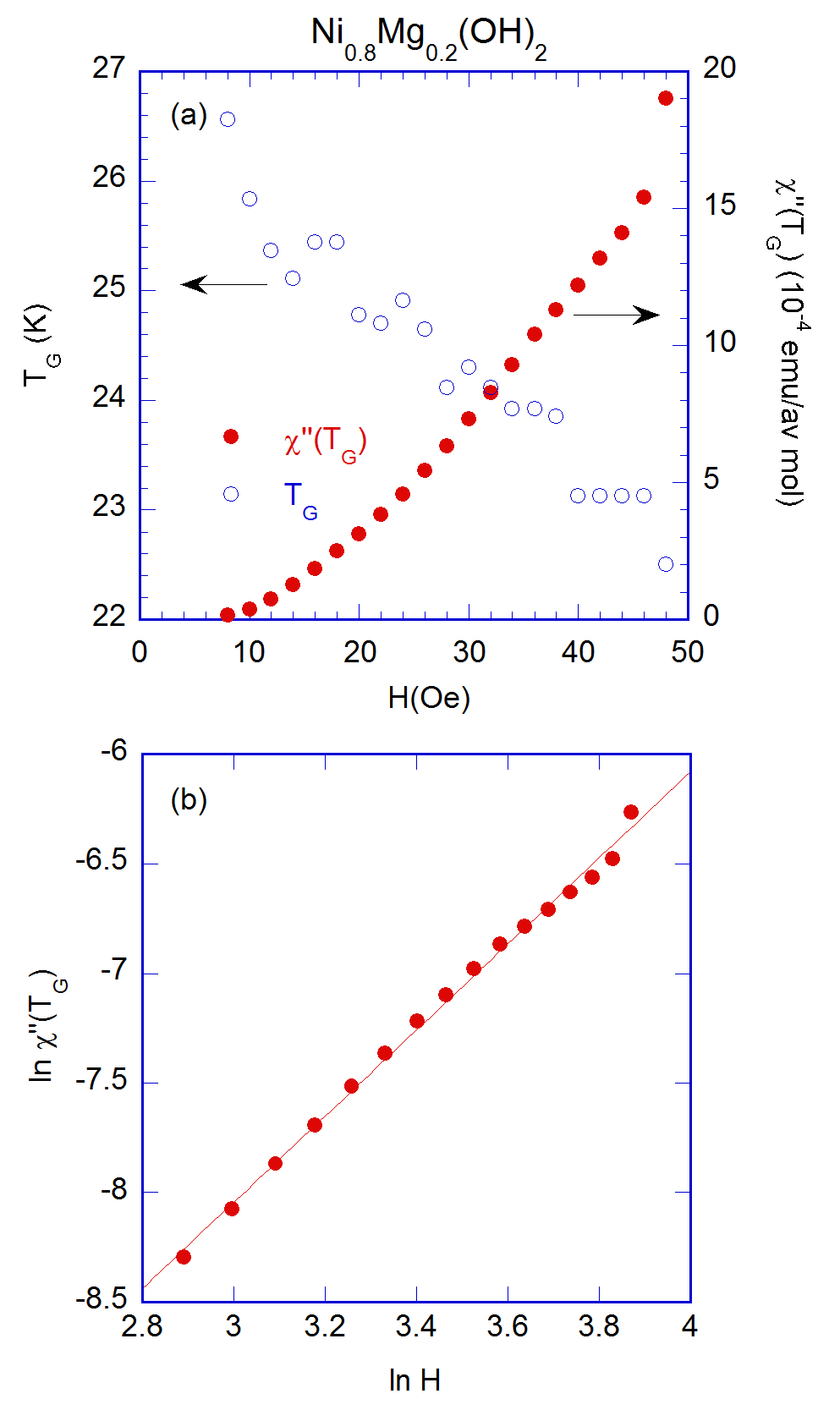}
\caption{\label{fig06}(Color online) (a) Plot of $T_{G}$ and the peak height $\chi^{\prime\prime}(T_{G})$ as a function of $H$ for $p=0.80$. (b) Plot of $\ln \chi^{\prime\prime}(T_{G})$ vs $\ln H$ for $18\le H\le 48$ kOe, where $H$ is in the units of kOe and $\chi^{\prime\prime}(T_{G})$ is the units of emu/av mol. The least-squares fitting curve is denoted by a straight line.}
\end{figure}

Figure \ref{fig05} shows the $T$ dependence of the absorption $\chi^{\prime\prime}$ for $p = 0.80$ at various magnetic fields, where $f=1$ Hz. A peak due to the metamagnetic transition drastically shifts to the low temperature side with increasing $H$. With increasing $H$, a broad peak of $\chi^{\prime\prime}$ vs $T$ associated with the Griffiths phase starts to appear well above $T_{N}$. This peak slightly shifts to the low temperature side with increasing $H$, while the peak height drastically increases. The separation between the peak due to the metamagnetic transition and the peak due to the appearance of the Griffiths phase, becomes more prominent for $H>20$ kOe. The peak due to the metamagnetic transition may disappear for $H>H_{c}$, where $H_{c} = 44 \pm 1$ kOe at $T = 1.56$ K (Enoki et al.\cite{ref12}).

Figure \ref{fig06}(a) shows the $H$ dependence of $T_{G}(H)$ for $p = 0.80$. This Griffiths temperature tends to decrease with increasing $H$, for example, $T_{G}$($H$ = 20 kOe) = 24.78 K and $T_{G}$($H$ = 48 kOe) = 22.50 K, reflecting the nature of the antiferromagnetic fluctuations (See Sec.~\ref{dis} for detail). Note that $T_{G}(H)$ is relatively higher than the N\'{e}el temperature $T_{N}$ = 20.7 K at $H=0$ for $p=0.80$. In Fig.~\ref{fig06}(a) we also show the $H$ dependence of the peak height $\chi^{\prime\prime}(T_{G})$ for $p=0.80$. The peak height drastically increases with increasing $H$ above 20 kOe. The least squares fit of the plot $\ln \chi^{\prime\prime}(T_{G})$ vs $\ln H$ (see Fig.~\ref{fig06}(b)) to Eq.(\ref{eq02}) for $18\le H\le 48$ kOe, yields the values of $A$ and $\zeta$ as 
\[
A = -13.93 \pm 0.11 ,\;\;\; \zeta = 1.96 \pm 0.03.
\]
The exponent $m$ for $M_{wall}$ is given 
by,
\[
m = \zeta + 1 = 2.96 \pm 0.03.
\]

\subsection{\label{resultC}Absorption $\chi^{\prime\prime}$ at $p=0.50$ in the presence of magnetic field}

\begin{figure}
\includegraphics[width=7.0cm]{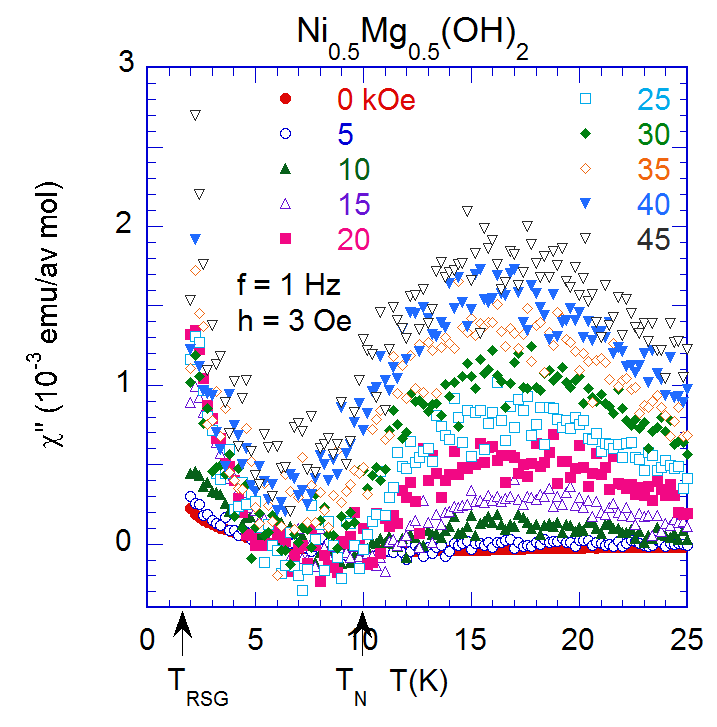}
\caption{\label{fig07}(Color online) $T$ dependence of $\chi^{\prime\prime}$ for $p = 0.50$ at various magnetic fields. $H$ = 0, 5, 10, 15, 20, 25, 30, 35, 40, and 45 kOe. $T_{RSG} = 5.3$ K and $T_{N}(H=0) = 10.1$ K\cite{ref14} (denoted by arrows). The critical field is $H_{c}=20 \pm 5$ kOe at $T = 1.56$ K (Enoki et al.\cite{ref12}). $f = 1$ Hz. $h = 3$ Oe. Each measurement of $\chi^{\prime\prime}$ vs $T$ was carried out in the FC state.}
\end{figure}

\begin{figure}
\includegraphics[width=7.0cm]{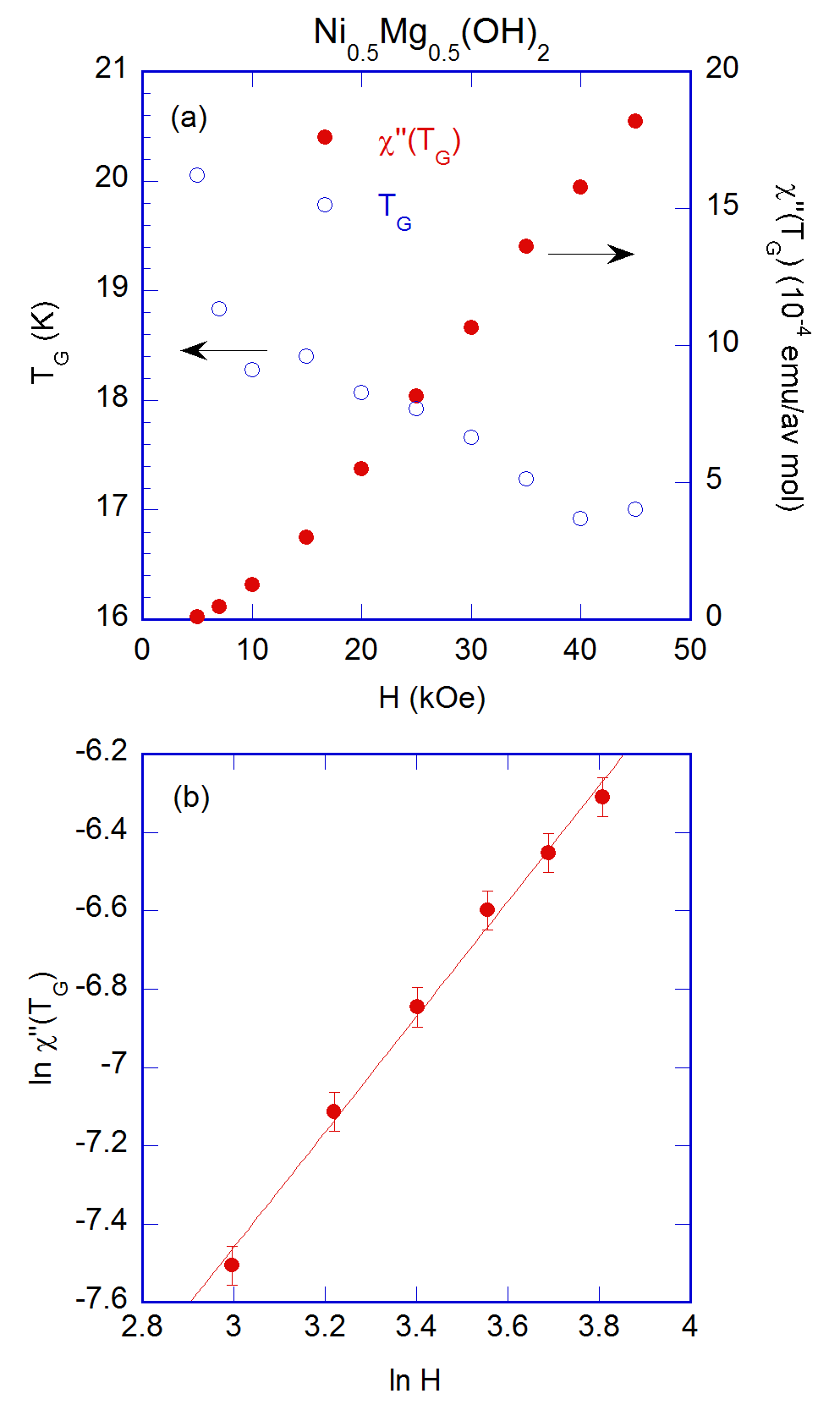}
\caption{\label{fig08}(Color online) (a) Plot of $T_{G}$ and the peak height $\chi^{\prime\prime}(T_{G})$ as a function of $H$ for $p = 0.50$. (b) Plot of $\ln \chi^{\prime\prime}(T_{G})$ vs $\ln H$ for $20\le H\le 45$ kOe, where $H$ is in the units of kOe and $\chi^{\prime\prime}(T_{G})$ is the units of emu/av mol. The least-squares fitting curve is denoted by a straight line.}
\end{figure}

Figure \ref{fig07} shows the $T$ dependence of the absorption $\chi^{\prime\prime}$ for $p = 0.50$ at various magnetic fields, where $f = 1$ Hz. A broad peak of $\chi^{\prime\prime}$ vs $T$ associated with the Griffiths phase starts to appear above $T_{N}$ (= 10.1 K) for $H>20$ kOe. The peak due to the metamagnetic transition may disappear for $H>H_{c}$, where $H_{c} = 20 \pm 5$ kOe at $T = 1.56$ K (Enoki et al.). 

Figure \ref{fig08}(a) shows the $H$ dependence of $T_{G}(H)$ for $p = 0.50$. This Griffiths temperature tends to decrease with increasing $H$, for example, $T_{G}$($H$ = 20 kOe) = 18.07 K and $T_{G}$($H$ = 45 kOe) = 17.00 K, reflecting the nature of the antiferromagnetic fluctuations. Note that $T_{G}(H)$ is relatively higher than the N\'{e}el temperature $T_{N} = 10.1$ K at $H=0$ for $p=0.50$. In Fig.~\ref{fig08}(a) we also show the $H$ dependence of the peak height $\chi^{\prime\prime}(T_{G})$ for $p=0.50$. This peak height drastically increases with increasing $H$ above 20 kOe. The least squares fit of the plot $\ln \chi^{\prime\prime}(T_{G})$ vs $\ln H$ (see Fig.~\ref{fig08}(b)) to Eq.(\ref{eq02}) for $20\le H\le 45$ kOe, yields the values of $A$ and $\zeta$ as 
\[
A = -11.88 \pm 0.20,\;\;\; \zeta = 1.47 \pm 0.06.
\]
The exponent $m$ for $M_{wall}$ is given by
\[
m = \zeta + 1 = 2.47 \pm 0.06.
\]

\subsection{\label{resultD}Absorption $\chi^{\prime\prime}$ at $p=0.315$ in the presence of magnetic field}

\begin{figure}
\includegraphics[width=7.0cm]{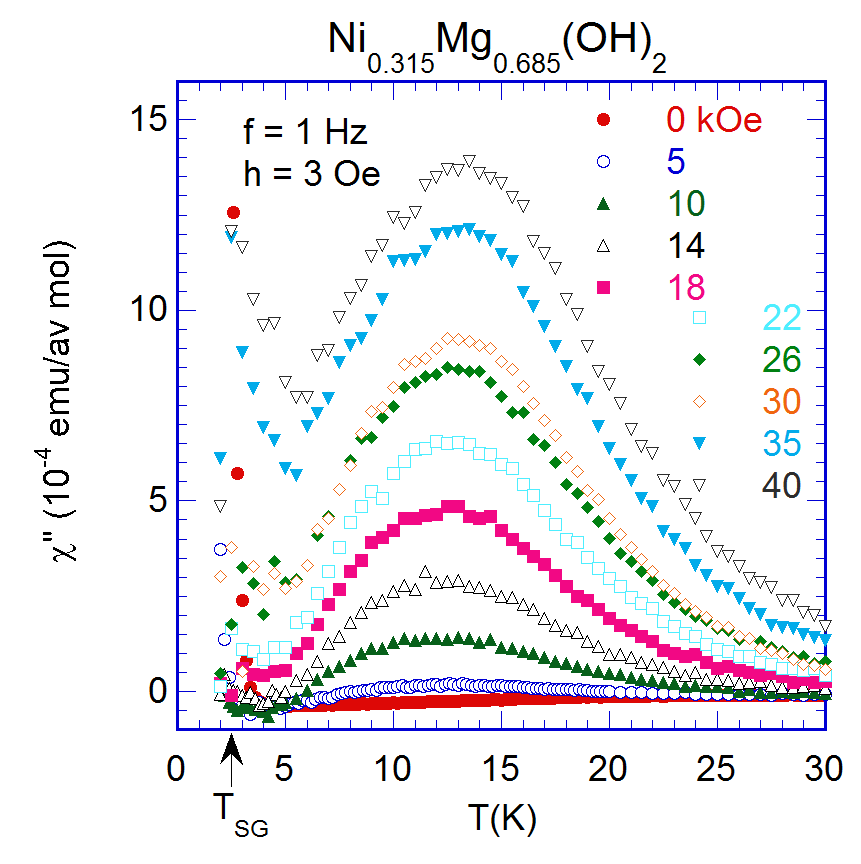}
\caption{\label{fig09}(Color online) $T$ dependence of $\chi^{\prime\prime}$ for $p=0.315$ at various magnetic fields. $H$ = 0, 5, 10, 14, 18, 22, 26, 30, 35, and 40 kOe. $T_{SG} = 2.5$ K\cite{ref14} (denoted by arrow). $f=1$ Hz. $h=3$ Oe. Each measurement of $\chi^{\prime\prime}$ vs $T$ was carried out in the FC state.}
\end{figure}

\begin{figure}
\includegraphics[width=7.0cm]{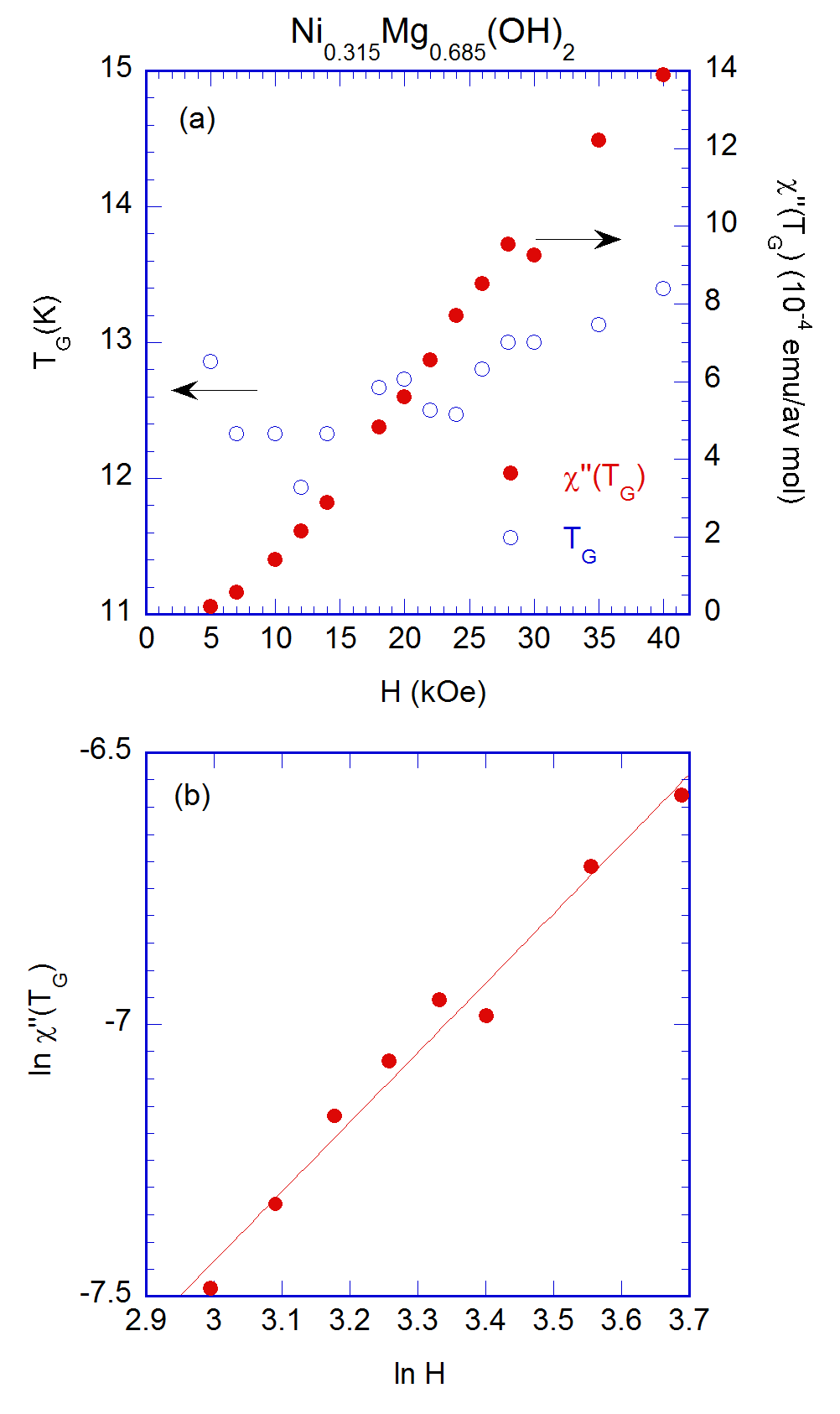}
\caption{\label{fig10}(Color online) (a) Plot of $T_{G}$ and the peak height $\chi^{\prime\prime}$ ($T=T_{G}$) as a function of $H$ for $p=0.315$. (b) Plot of $\ln \chi^{\prime\prime}(T_{G})$ vs $\ln H$ for $20\le H\le 40$ kOe, where $H$ is in the units of kOe and $\chi^{\prime\prime}(T_{G})$ is the units of emu/av mol. The least-squares fitting curve is denoted by a straight line.}
\end{figure}

Figure \ref{fig09} shows the $T$ dependence of the absorption $\chi^{\prime\prime}$ for $p=0.315$ at various magnetic fields, where $f = 1$ Hz. A broad peak of $\chi^{\prime\prime}$ vs $T$ associated with the Griffiths phase starts to appear above the spin glass transition temperature $T_{SG}$ (= 2.5 K), for $H>20$ kOe. 

Figure \ref{fig10}(a) shows the $H$ dependence of $T_{G}(H)$ for $p=0.315$. The Griffiths temperature tends to increase with increasing $H$, for example, $T_{G}$($H$ = 20 kOe) = 12.73 K and $T_{G}$($H$ = 40 kOe) = 13.4 K. Note that $T_{G}(H)$ is much higher than $T_{SG}$ (= 2.5 K). This increase in $T_{G}(H)$ with $H$ for $p=0.315$ is in contrast with the decrease of $T_{G}(H)$ with $H$ for $p=0.50$. This increase of $T_{G}(H)$ may suggest the nature of the ferromagnetic fluctuations for $p=0.315$. In Fig.~\ref{fig10}(a) we also show the $H$ dependence of the peak height $\chi^{\prime\prime}(T_{G})$ for $p=0.315$. This peak height drastically increases with increasing $H$ above 20 kOe. The least squares fit of the plot $\ln \chi^{\prime\prime}(T_{G})$ vs $\ln H$ (see Fig.~\ref{fig10}(b)) to Eq.(\ref{eq02}) for $20\le H\le 40$ kOe, yields the values of $A$ and $\zeta$ as 
\[
A = -11.27 \pm 0.25,\;\;\; \zeta = 1.28 \pm 0.08.
\]
The exponent $m$ for $M_{wall}$ is given by, 
\[
m = 2.28 \pm 0.08.
\]

\subsection{\label{resultE}Absorption $\chi^{\prime\prime}$ at $p=0.25$ in the presence of magnetic field}

\begin{figure}
\includegraphics[width=7.0cm]{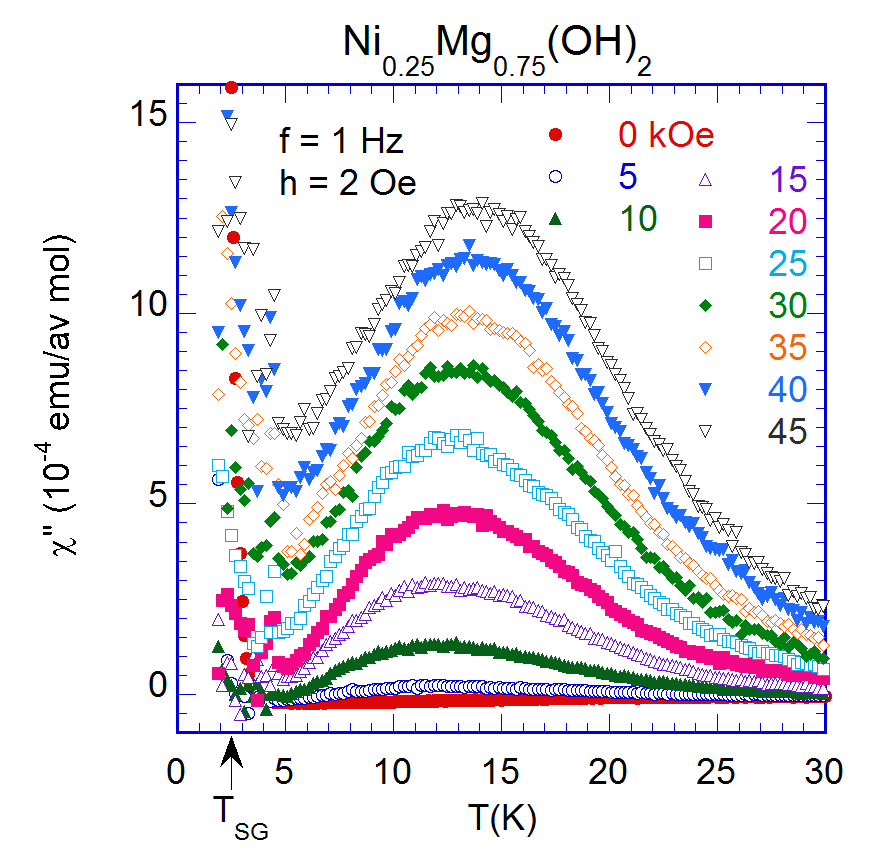}
\caption{\label{fig11}(Color online) $T$ dependence of $\chi^{\prime\prime}$ for $p=0.25$ at various magnetic fields. $H$ = 0, 5, 10, 15, 20, 25, 30, 35, 40, and 45 kOe. $T_{SG}=2.5$ K\cite{ref14} (denoted by arrow). $f=1$ Hz. $h=2$ Oe. Each measurement of $\chi^{\prime\prime}$ vs $T$ was carried out in the FC state.}
\end{figure}

\begin{figure}
\includegraphics[width=7.0cm]{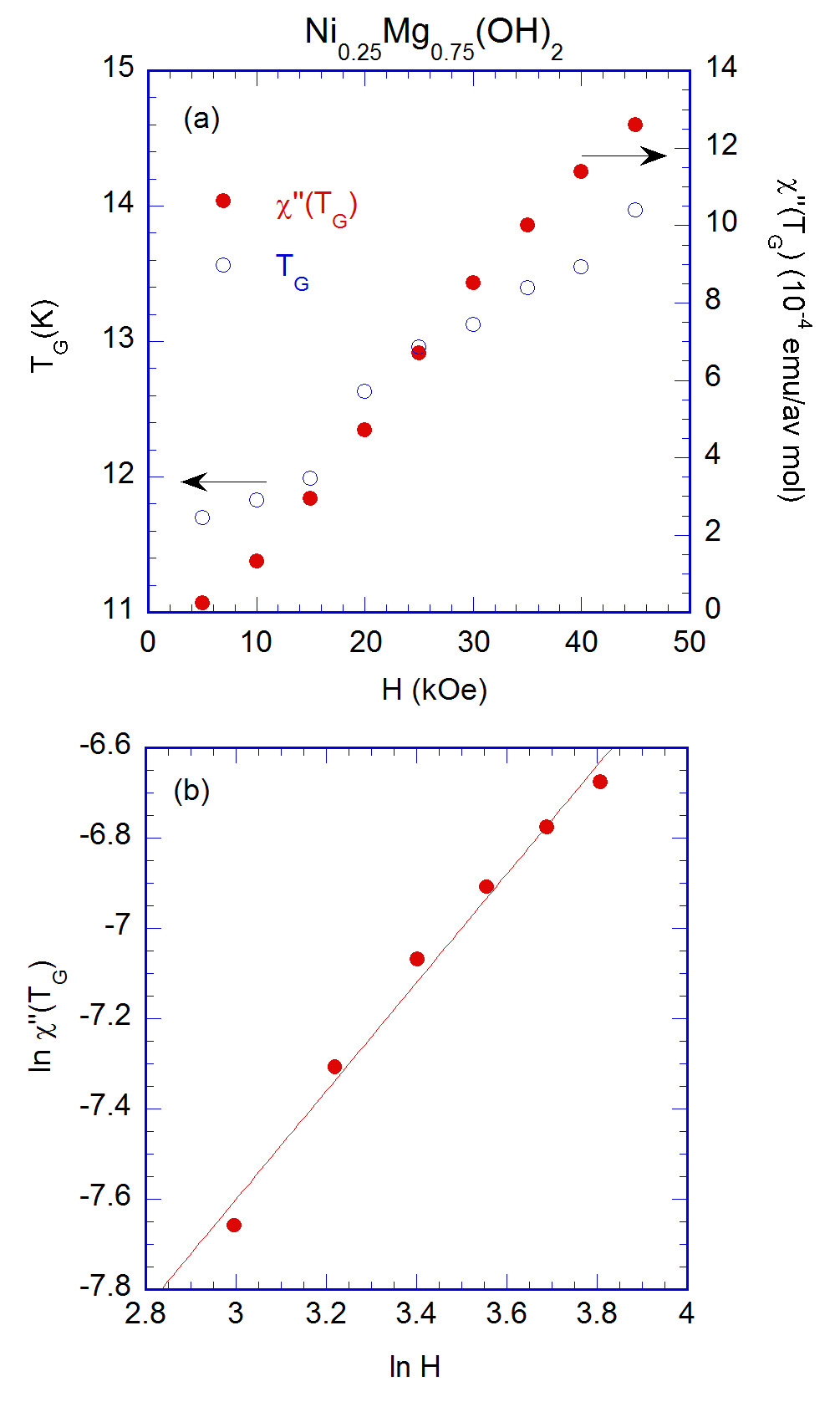}
\caption{\label{fig12}(Color online) (a) Plot of $T_{G}$ and the peak height $\chi^{\prime\prime}$ ($T=T_{G}$) as a function of $H$ for $p=0.25$. (b) Plot of $\ln \chi^{\prime\prime}(T_{G})$ vs $\ln H$ for $20\le H\le 5$ kOe, where $H$ is in the units of kOe and $\chi^{\prime\prime}(T_{G})$ is the units of emu/av mol. The least-squares fitting curve is denoted by a straight line.}
\end{figure}

Figure \ref{fig11} shows the $T$ dependence of the absorption $\chi^{\prime\prime}$ for $p=0.25$ at various magnetic fields, where $f = 1$ Hz. A broad peak of $\chi^{\prime\prime}$ vs $T$ associated with the Griffiths phase starts to appear above the spin glass transition temperature $T_{SG}$ (= 2.5 K), for $H>20$ kOe. Figure \ref{fig12}(a) shows the $H$ dependence of $T_{G}(H)$ for $p=0.25$. The increase of $T_{G}(H)$ increase with increasing $H$, for example, $T_{G}$($H$ = 20 kOe) = 12.63 K and $T_{G}$($H$ = 45 kOe) = 13.97 K, reflects the nature of the ferromagnetic fluctuations. In Fig.~\ref{fig12}(a) we also show the $H$ dependence of the peak height $\chi^{\prime\prime}(T_{G})$ for $p=0.25$. This peak height drastically increases with increasing $H$ above 20 kOe. The least squares fit of the plot $\ln \chi^{\prime\prime}(T_{G})$ vs $\ln H$ (see Fig.~\ref{fig12}(b)) to Eq.(\ref{eq02}) for $20<H<45$ kOe, yields the values of $A$ and $\zeta$ as 
\[
A = -11.21 \pm 0.24,\;\;\; \zeta = 1.20 \pm 0.07.
\]
The exponent $m$ for $M_{wall}$ is given by 
\[
m = 2.20 \pm 0.07.
\]

\subsection{\label{resultF}Absorption $\chi^{\prime\prime}$ at $p=0.10$ in the presence of magnetic field}

\begin{figure}
\includegraphics[width=7.0cm]{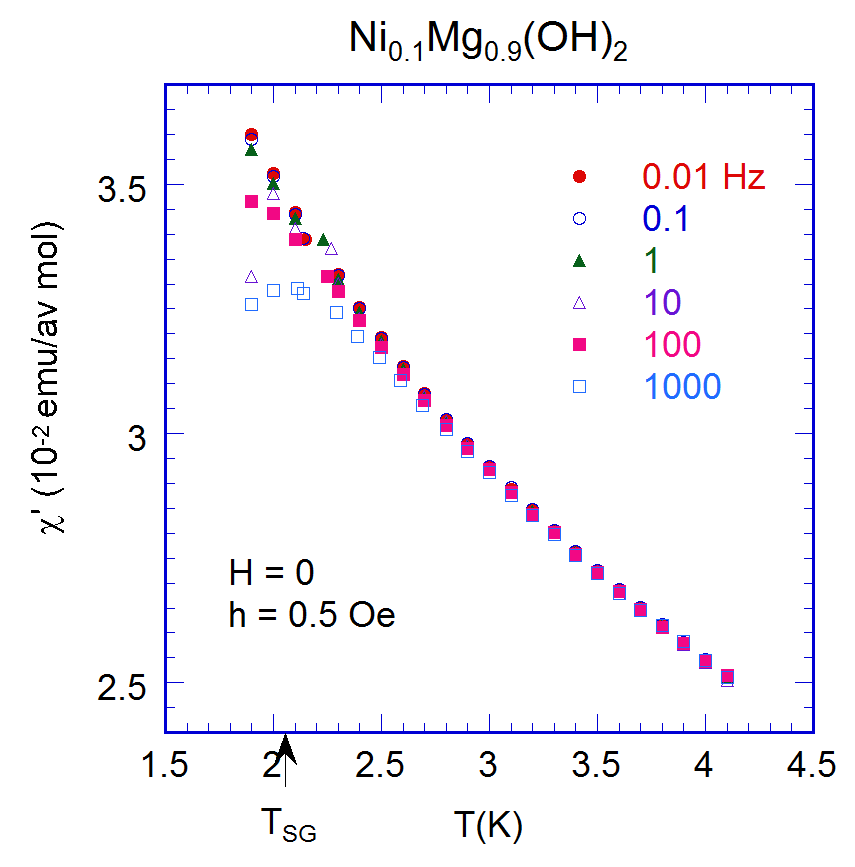}
\caption{\label{fig13}(Color online) $T$ dependence of $\chi^{\prime}$ for $p=0.10$ at various frequency. $H=0$. $h=0.5$ Oe.}
\end{figure}

\begin{figure}
\includegraphics[width=7.0cm]{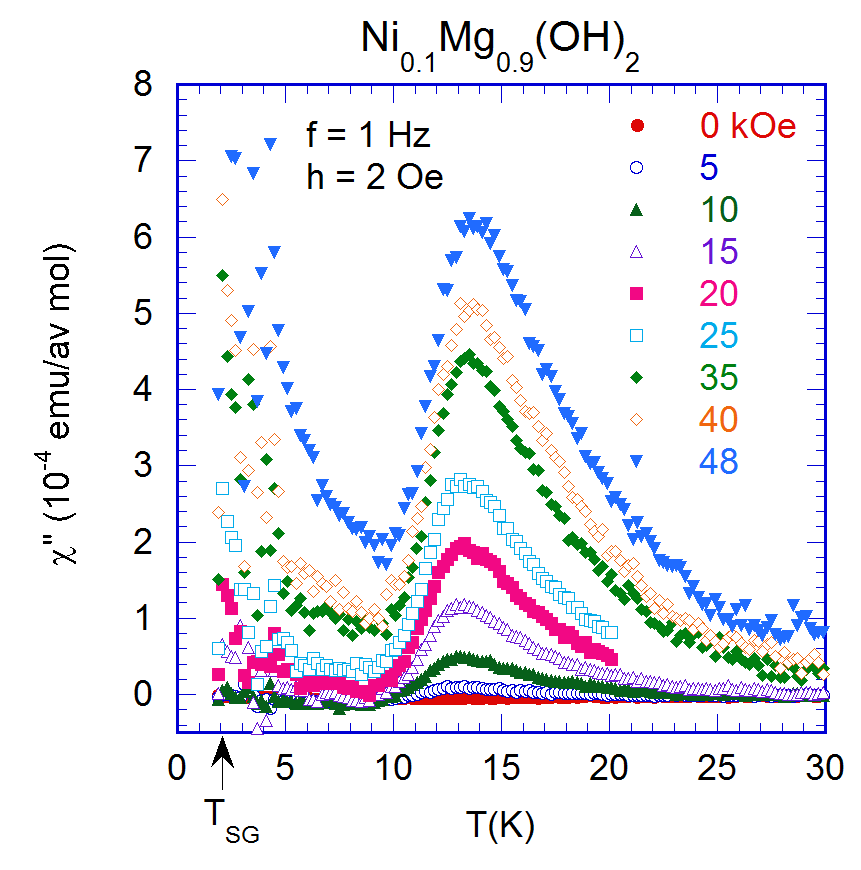}
\caption{\label{fig14}(Color online) $T$ dependence of $\chi^{\prime\prime}$ for $p=0.10$ at various magnetic fields. $H$ = 0, 5, 10, 15, 20, 25, 30, 35, 40, and 48 kOe. $T_{SG}=2.05$ K\cite{ref14} (denoted by arrow). $f=1$ Hz. $h=2$ Oe. Each measurement of $\chi^{\prime\prime}$ vs $T$ was carried out in the FC state.}
\end{figure}

\begin{figure}
\includegraphics[width=7.0cm]{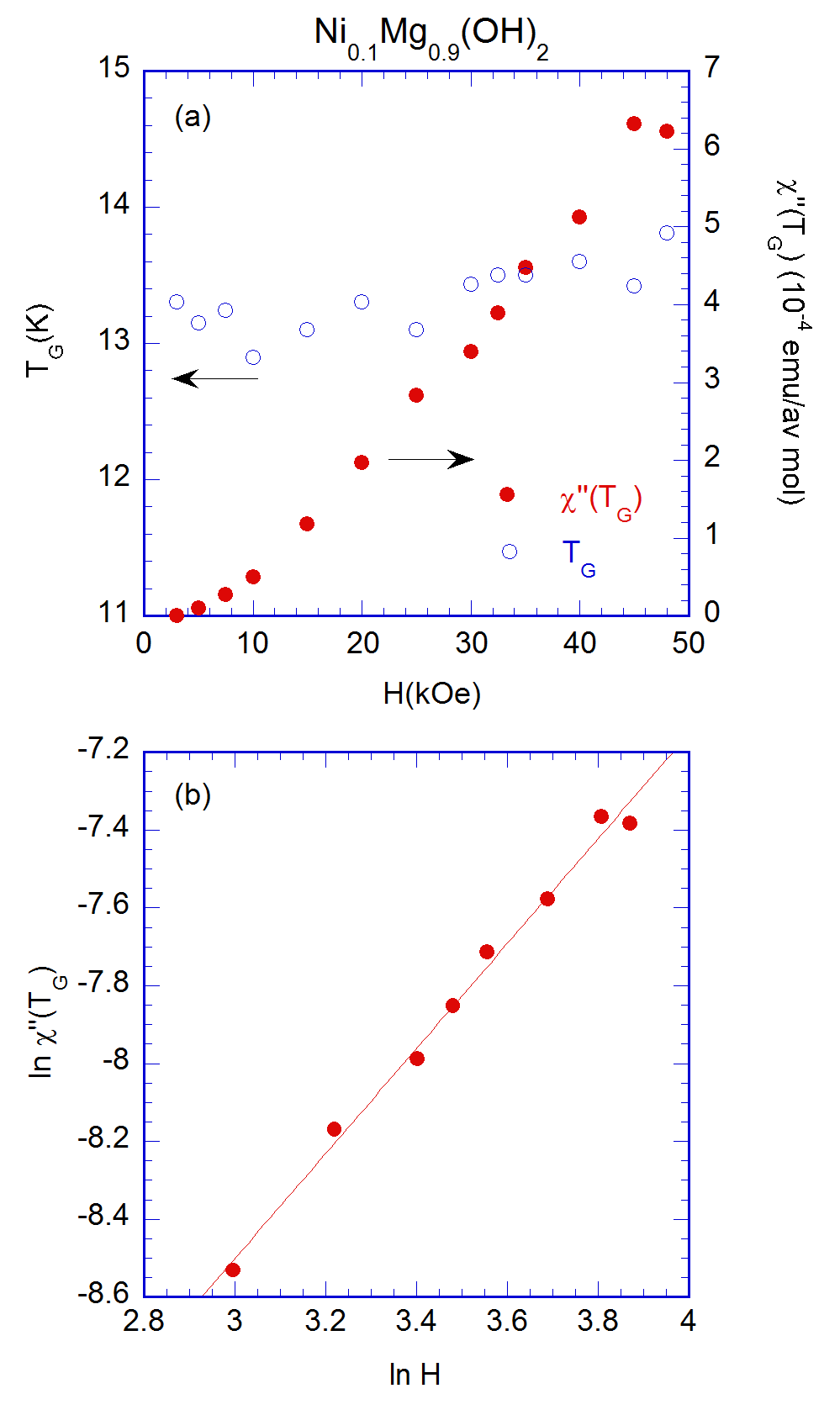}
\caption{\label{fig15}(Color online) (a) Plot of $T_{G}$ and the peak height $\chi^{\prime\prime}$ ($T=T_{G}$) as a function of $H$ for $p=0.10$. (b) Plot of $\ln \chi^{\prime\prime}(T_{G})$ vs $\ln H$ for $20\le H\le 45$ kOe, where $H$ is in the units of kOe and $\chi^{\prime\prime}(T_{G})$ is the units of emu/av mol.}
\end{figure}

Figure \ref{fig13} shows the $T$ dependence of $\chi^{\prime}$ for $p=0.10$ at $H=0$ was also measured at various $f$: $0.01 \le f \le 1000$ Hz, where $h=0.5$ Oe. The dispersion $\chi^{\prime}$ increases with decreasing T for $0.01 \le f \le 200$ Hz. It exhibits a broad peak around 2.05 K at $f = 330$ Hz, which slightly shifts to the high temperature side with increasing $f$. In contrast, $\chi^{\prime\prime}$ starts to appear below 3 K and increases with decreasing $T$. The increase of $\chi^{\prime\prime}$ with decreasing $T$ below 2.1 K becomes more remarkable as $f$ increases.

Figure \ref{fig14} shows the $T$ dependence of the absorption $\chi^{\prime\prime}$ for $p=0.10$ at various magnetic fields, where $f=1$ Hz. A broad peak of $\chi^{\prime\prime}$ vs $T$ associated with the Griffiths phase starts to appear above the spin glass transition temperature $T_{SG}$ (= 2.05 K), for $H>20$ kOe. Figure \ref{fig15}(a) shows the $H$ dependence of $T_{G}(H)$ for $p = 0.10$. The slight increase of $T_{G}(H)$ with increasing $H$, for example, $T_{G}$($H$ = 20 kOe) = 13.3 K and $T_{G}$($H$ = 48 kOe) = 13.81 K, reflects the nature of the ferromagnetic fluctuations. In Fig.~\ref{fig15}(a) we also show the $H$ dependence of the peak height $\chi^{\prime\prime}(T_{G})$ for $p=0.10$. This peak height drastically increases with increasing $H$ above 20 kOe. The least squares fit of the plot $\ln \chi^{\prime\prime}(T_{G})$ vs $\ln H$ (see Fig.~\ref{fig15}(b)) to Eq.(\ref{eq02}) for $20\le H\le 45$ kOe, yields the values of $A$ and $\zeta$ as 
\[
A = -12.55 \pm 0.18,\;\;\; \zeta = 1.35 \pm 0.05 .
\]
The exponent $m$ for $M_{wall}$ is given by
\[
m = 2.35 \pm 0.07.
\]

\section{\label{dis}DISCUSSION}
\subsection{\label{disA}Domain wall magnetization}
Here we discuss the physical meaning of the exponent $m$ for the domain-wall magnetization $M_{wall}$.\cite{ref17,ref18} The principal contribution at higher fields is due to the domain walls where parallel spins preferentially align with the field thus enhancing the magnetization. The bulk contribution is normally small and originates from a statistical imbalance between up and down spins within the domains. If $R$ is the characteristic domain size, the wall area grows as $R^{2}$ as the domains grow. Because the total number of spins is fixed, however, the number of domains decreases as $R^{-3}$. Thus the total number of spins in the walls decreases as $R^{-1}$. The magnetization from the domain walls in the FC state can be related to the domain size $R$, 
\begin{equation} 
M_{wall}\approx R^{-1}.
\label{eq04} 
\end{equation} 
The field dependence of $M_{wall}$ may be described by
\begin{equation}
M_{wall} \approx H^{m} ,
\label{eq05} 
\end{equation} 
at low temperatures, where $R\approx H^{-m}$ and $m$ is the exponent for the domain size. Here we assume that the absorption $\chi^{\prime\prime}$ may be given by
\begin{equation}
\chi^{\prime\prime}\approx \frac{M_{wall}}{H} \approx H^{m-1} \approx H^{\zeta} ,
\label{eq06} 
\end{equation} 
with
\[
m = \zeta + 1.
\]
Since the exponent $\zeta$ can be determined experimentally, the exponent $m$ can be obtained for each Ni concentration $p$. 

\begin{figure}
\includegraphics[width=7.0cm]{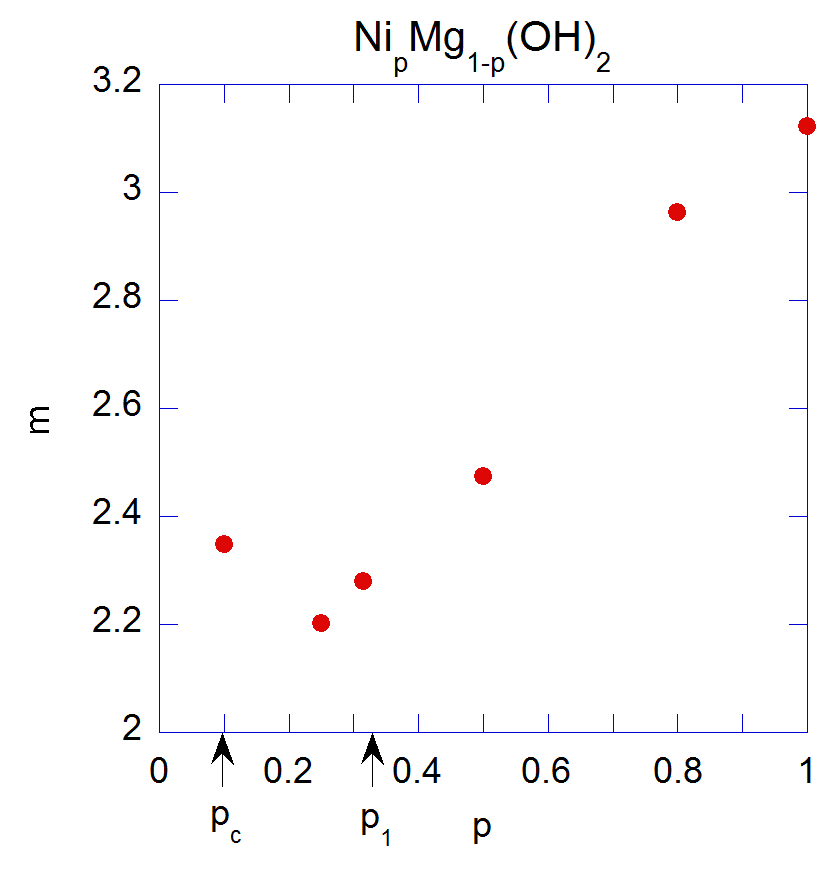}
\caption{\label{fig16}(Color online) Exponent $m$ vs Ni concentration $p$ for Ni$_{p}$Mg$_{1-p}$(OH)$_{2}$. $p_{c}$ (= 0.1) and $p_{1}$ (= 1/3) (denoted by arrows).}
\end{figure}

Figure \ref{fig16} shows the exponent $m$ as a function of Ni concentration for Ni$_{p}$Mg$_{1-p}$(OH)$_{2}$. The exponent $m$ is determined as $3.12 \pm 0.04$ for $p=1$, $2.96 \pm 0.03$ for $p=0.80$, $2.47 \pm 0.06$ for $p=0.50$, $2.28 \pm 0.08$ for $p=0.315$, $2.20 \pm 0.07$ for $p=0.25$, and $2.35 \pm 0.07$ for $p=0.10$. It is found that $m$ decreases with decreasing Ni concentration and takes a minimum around $p\approx p_{1}=0.33$, where $p_{1}$ is the estimated critical concentration when only the first neighbor interaction is taken into account; $p_{1}=2/z_{1}=1/3$. Our results of $m$ are on the same order as those for other dilute Ising random antiferromagnets; $m = 2.9 \pm 0.1$ for Fe$_{0.6}$Zn$_{0.4}$F$_{2}$ (Mattsson et al.\cite{ref18}), $m = 3.2 \pm 0.3$ for Fe$_{0.46}$Zn$_{0.56}$F$_{2}$ (Ledermann et al.\cite{ref17}), and $m = 2.7 \pm 0.1$ for Fe$_{0.5}$Zn$_{0.5}$F$_{2}$ (Feng et al.\cite{ref19}). According to Mattsson et al.,\cite{ref18} $m$ can be expressed in terms of 
\begin{equation}
m = \nu_{H}+(2D_{s}+2-2d)/(d-2),
\label{eq07} 
\end{equation} 
where $d$ is the spatial dimensionality of the system, $\nu_{H}$ is the exponent for the system with smooth domain walls, and $D_{s}$ is the fractal dimension of the domain walls on short length scales. For $d = 3$, we have 
\begin{equation}
m = \nu_{H} + (2D_{s}-4).
\label{eq08} 
\end{equation} 
The exponent $\nu_{H}$ can be evaluated from the neutron scattering experiment. If we assume that $\nu_{H}$ (= 2.1)\cite{ref18} may be independent of the Ni concentration $p$, the fractal dimension $D_{s}$ can be evaluated as $D_{s}=2.5$ for $m=3.12$ ($p=1$) and $D_{s}=2.1$ for $m = 2.28$ ($p=0.315$). This means that the fractal dimension decreases from 2.5 to 2.1 with the dilution of Mg$^{2+}$ ions. These value of $D_{s}$ may be consistent with the theoretical predictions; $D_{s} = 2.27 \pm 0.02$ by Middleton and Fisher.\cite{ref20} 

\subsection{\label{disB}Griffiths phase in the magnetic phase diagram}

\begin{figure}
\includegraphics[width=7.0cm]{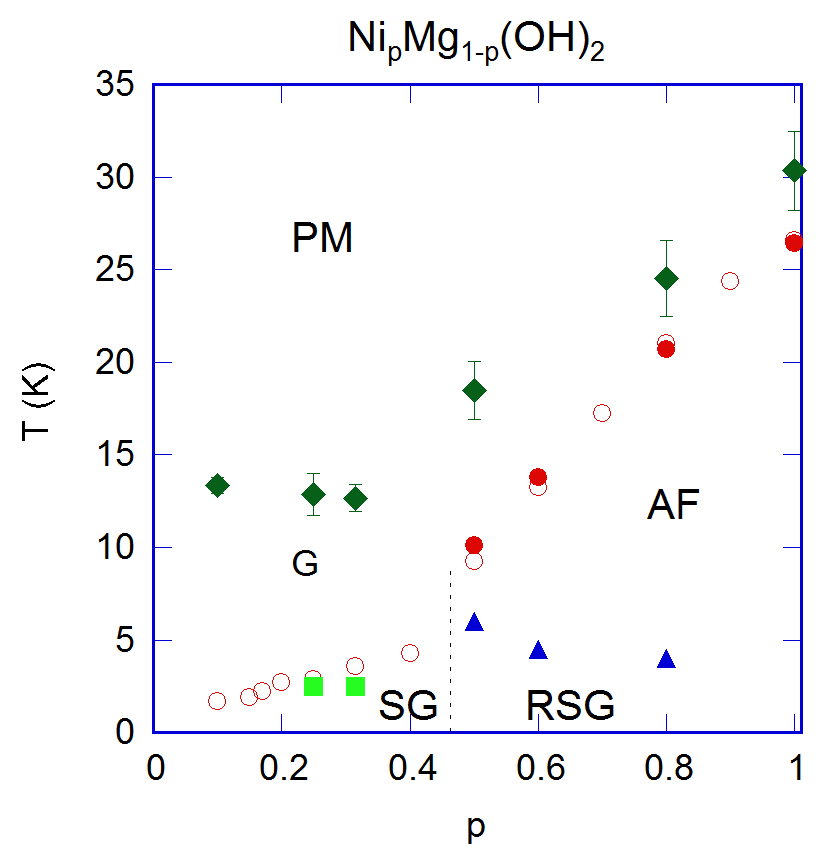}
\caption{\label{fig17}(Color online) Magnetic phase diagram (critical temperature vs Ni concentration $p$) of Ni$_{p}$Mg$_{1-p}$(OH)$_{2}$. PM: paramagnetic phase, AF: antiferromagnetic phase, SG: spin-glass phase, RSG: reentrant spin-glass phase, and G: Griffiths phase. The original magnetic phase diagram was reported by Suzuki et al.\cite{ref14} (see the detail for the definition of open circle, closed square, closed triangle in the Reference\cite{ref14}). The magnetic-field-induced Griffiths phase is plotted in this original magnetic phase diagram. Note that the Griffith temperature $T_{G}$ (closed diamond) is a peak temperature where the absorption $\chi^{\prime\prime}$ exhibits a peak in the presence of $H$ (typically $20\le H\le 50$ kOe). The bars of the data of $T_{G}$ denotes the change of $T_{G}$ as $H$ is changed. In this sense, the Griffiths phase is the magnetic-induced phase. The data reported by Enoki and Tsujikawa\cite{ref09} are denoted by open circles. The dotted vertical line ($p\simeq 0.46$) is the phase boundary between the RSG phase and the SG phase.}
\end{figure}

Figure \ref{fig17} shows the magnetic phase diagram for Ni$_{p}$Mg$_{1-p}$(OH)$_{2}$. The original magnetic phase diagram was reported by Suzuki et al.\cite{ref14} The magnetic-field-induced Griffiths phase is plotted in this original magnetic phase diagram. Here the Griffith temperature $T_{G}$ is defined by a peak temperature where the absorption $\chi^{\prime\prime}$ exhibits a peak in the presence of $H$ (typically $20\le H\le 50$ kOe). It is found that $T_{G}$ is strongly dependent on the Ni concentration $p$ in our system. This result is rather different from our expectation before the measurements: $T_{G}$ is independent of the concentration and coincides with the critical temperature of the undiluted system ($p=1$). For a ferromagnet with site or bond dilution, it is predicted that the Griffiths phase G is the region between the horizontal line ($T=T_{G}$) and the phase boundary for the onset of ferromagnetism. The latter boundary meets the zero-temperature axis at the percolation threshold $p_{c}$. The Griffiths temperature $T_{G}$ is the critical temperature of the undiluted system ($p=1$).\cite{ref21} 

For $p$ = 1, 0.60, and 0.50, the Griffiths temperature $T_{G}$ is higher than the corresponding N\'{e}el temperature. It decreases with increasing $H$ for $H>20$ kOe. This behavior may be due to the antiferromagnetic nature of the spin fluctuations. The effective interplanar antiferromagnetic interactions is expressed by $J^{\prime}(\xi_{a}/a)^{2}$, where $a$ is the in-plane lattice constant, $\xi_{a}$ is the in-plane ferromagnetic spin correlation length, and $J^{\prime}$ is the antiferromagnetic interplanar interaction such as $J_{2}$ and $J_{3}$ in our system. This effective interplanar antiferromagnetic interaction becomes dominant near the N\'{e}el temperature through the dramatic growth of the in-plane ferromagnetic spin correlation length. 

For $p$ = 0.315, 0.25, and 0.10, $T_{G}$ is higher than the spin glass freezing temperature $T_{SG}$. It is almost independent of $p$: $T_{G} \approx 13.7$ K. The Griffiths temperature slightly increases with increasing $H$. This behavior may be due to the ferromagnetic nature of the spin fluctuation. The effective interplanar antiferromagnetic interactions are much weaker than the intraplanar exchange interaction since there is only ferromagnetic short range order at any finite temperature below the percolation threshold ($p=p_{1}$). 

Finally we note that the existence of Griffiths phase has been confirmed by Deguchi et al.\cite{ref16} in Ni$_{p}$Mg$_{1-p}$(OH)$_{2}$ with $p=0.42$ from the slow decay of thermoremnant magnetization $M_{TRM}$. The system is quenched from 50 K to a temperature in the presence of magnetic field $H$ (= 100 Oe) (FC cooling) and annealed at a temperature $T_{0}$ for a wait time $t_{w}$ ( = 10 min). After the magnetic field $H$ is reduced to zero at $t=0$, $M_{TRM}$ is measured as a function of time. The slow relaxation of $M_{TRM}$ is clearly observed in the Griffiths phase ($T_{N}<T_{0}<T_{G}$); where $T_{N} \approx 7.5$ K and $T_{G}$ (= 17 K). This value of $T_{G}$ for $p=0.42$ is a little lower than the value of $T_{G}$ (= 14 K) evaluated from our magnetic phase diagram. 

\section{CONCLUSION}
The nature of the magnetic-field-induced Griffiths phase in 3D Ising random magnet Ni$_{p}$Mg$_{1-p}$(OH)$_{2}$ ($p$ = 0.10, 0.25, 0.315, 0.50, 0.80, and 1) has been studied from the absorption $\chi^{\prime\prime}$ (the out-of phase in AC magnetic susceptibility) in the field-cooled (FC) state. The Griffiths temperature $T_{G}$ is defined as the peak temperature of $\chi^{\prime\prime}$ vs $T$ above $T_{N}(H=0)$. The peak height $\chi^{\prime\prime}(T_{G})$ drastically increases with increasing $H$ for $H>20$ kOe, following a power form, $\chi^{\prime\prime}(T_{G}) \approx H^{m-1}$. The exponent $m$ depends on the Ni concentration; $m = 3.12 \pm 0.04$ for $p=1$ and $2.28 \pm 0.08$ for $p=0.315$. 

\begin{acknowledgments}
We are grateful to Prof. Toshiaki Enoki for providing us with samples of Ni$_{p}$Mg$_{1-p}$(OH)$_{2}$ and for his invaluable suggestions and discussions. 
\end{acknowledgments}

\end{document}